\documentclass[conference]{IEEEtran}
\newenvironment{aiprompt}{\begin{quote}}{\end{quote}}
\usepackage{cite}
\usepackage{balance}
\usepackage{amsmath,amssymb,amsfonts}
\usepackage{algorithmic}
\usepackage{graphicx}
\usepackage{textcomp}
\usepackage{xcolor}
\def\BibTeX{{\rm B\kern-.05em{\sc i\kern-.025em b}\kern-.08em
    T\kern-.1667em\lower.7ex\hbox{E}\kern-.125emX}}
\begin{document}

\title{Generative AI for \mbox{Multiple Choice STEM Assessments}\\
}

\author{\IEEEauthorblockN{%1\textsuperscript{st}
Christina Spirou Perdikoulias}
\IEEEauthorblockA{\textit{Digital Education Company Ltd.} \\
Waterloo, Ontario, Canada \\
cp@digitaled.com}
\and
\IEEEauthorblockN{%2\textsuperscript{nd}
Chad Vance}
\IEEEauthorblockA{\textit{Digital Education Company Ltd.} \\
Waterloo, Ontario, Canada \\
cvance@digitaled.com}
\and
\IEEEauthorblockN{%3\textsuperscript{rd}
Stephen M. Watt}
\IEEEauthorblockA{\textit{Cheriton School of Computer Science} \\
\textit{University of Waterloo}\\
Waterloo, Ontario, Canada \\
smwatt@uwaterloo.ca}
}

\maketitle

\begin{abstract}
Artificial intelligence (AI) technology enables a range of enhancements in computer-aided instruction, from accelerating the creation of teaching materials to customizing learning paths based on learner outcomes. However, ensuring the mathematical accuracy and semantic integrity of generative AI output remains a significant challenge, particularly in Science, Technology, Engineering and Mathematics (STEM) disciplines. In this study, we explore the use of generative AI in which ``hallucinations'', typically viewed as undesirable inaccuracies, can instead serve a pedagogical purpose. Specifically, we investigate the generation of plausible but incorrect alternatives for multiple choice assessments, where credible distractors are essential for effective assessment design. We describe the Moebius platform for online instruction, with particular focus on its architecture for handling mathematical elements through specialized semantic packages that support dynamic, parameterized STEM content. We examine methods for crafting prompts that interact effectively with these mathematical semantics to guide the AI in generating high-quality multiple choice distractors. Finally, we demonstrate how this approach reduces the time and effort associated with creating robust teaching materials while maintaining academic rigor and assessment validity.
\end{abstract}


\section{Introduction}
Online learning environments offer unique opportunities to enhance student outcomes, particularly in the context of STEM education. Dynamic, interactive content that encourages discovery and exploration has been shown to foster deeper learning compared to passive instructional approaches. We operate under the belief that the effectiveness of online STEM education is significantly enhanced when learners engage with content that promotes “desirable difficulties” — exercises that are intentionally challenging, allowing students to experience struggle, failure, and corrective feedback as part of the learning process. As~\cite{dunlosky2013} observes, “We are best positioned to optimize memory through ‘desirable difficulties.’ In this sense, learning is about challenge, failure, understanding why something is wrong before getting it right”. Immediate, targeted feedback within these exercises serves a critical role by redirecting the learner when misconceptions arise, thus supporting more durable understanding and retention.

The effectiveness of online STEM instruction is further amplified by personalized learning paths that adapt to each student’s evolving competency~\cite{Yousef2024}. Adaptive systems allow learners to progress efficiently through concepts they have already mastered, while providing additional practice and support in areas where their understanding is weaker. Personalized pathways help optimize the learner's engagement and ensure that instructional time is focused where it is most impactful.

A key enabler of dynamic, personalized online learning experiences is the ability to generate contextually relevant questions ``on-the-fly,'' as learners engage with instructional materials and advance through course concepts. Traditionally, this generation of context-constrained questions has been achieved through algorithmic methods supported by a competent computation engine. A powerful computation engine not only enables the dynamic creation of mathematically rigorous questions but also supports automatic grading, symbolic manipulation, and the evaluation of equivalency in student responses~\cite{Barana2021}. 

Recent advances in large language model (LLM) research suggest complementary approaches for validating and diversifying automatically generated educational content. Recent work by Alhazmi et al.~\cite{Alhazmi2024} provides a comprehensive survey of distractor generation methods, highlighting the evolution from rule-based approaches to large language model–driven techniques. Their analysis underscores the continuing challenge of balancing plausibility, semantic accuracy, and educational validity. Similarly, Feng et al.~\cite{feng-etal-2024-exploring} evaluated large language model approaches for generating distractors in mathematics multiple-choice questions, finding that in-context prompting produced mathematically valid alternatives comparable to those authored by humans. Their results highlight the promise of LLMs for efficient content creation while underscoring the need for validation mechanisms that capture authentic student misconceptions. Algorithms designed for question generation leverage these engines to produce parameterized problem sets, ensuring both variation and consistency, and to automate feedback processes that are central to adaptive learning environments.

Our experience in building question generation engines for STEM disciplines has reinforced the importance of minimizing the effort required to create and refine high-quality assessment content. This consideration is particularly urgent within the current landscape of higher education in North America, where institutions face increasingly constrained budgets and resource limitations. At the same time, the pace of technological advancement within STEM fields demands that course materials remain relevant and up to date, requiring instructors and course designers to iterate and refresh content multiple times each academic year. The ability to rapidly generate, adjust, and redeploy assessment materials is therefore essential to maintain the relevance, rigor, and responsiveness of STEM education programs.

\subsection{Challenges of Using Generative AI in Mathematical Contexts}
While generative AI has shown considerable promise in educational settings, its application within mathematics continues to present challenges. Mathematical reasoning relies heavily on precision, symbolic logic, and domain-specific conventions that are difficult for large language models (LLMs) to reliably reproduce. A growing body of research has documented the limitations of generative models when tasked with producing or evaluating mathematical expressions. These limitations are especially evident in multi-step reasoning problems, symbolic manipulation, and the preservation of mathematical definitions or formal constraints.

For instance,~\cite{Boye2025} conducted an evaluation of various advanced LLMs, including GPT-4o and Mixtral, using word problems typically encountered in Secondary-level mathematics. Their findings revealed that all exhibited errors of varying levels of accuracy, in spatial reasoning, strategic planning, and arithmetic. Notably, some models produced correct answers through incorrect logic, indicating a lack of genuine understanding in their problem-solving processes. ``Common failure modes include unwarranted assumptions, over-reliance on numerical patterns, and difficulty translating physical intuition into mathematical steps''~\cite{Boye2025}.

Another study investigated the capabilities of large pretrained language models in arithmetic and symbolic induction tasks. One study~\cite{Qian2022} found that these large pretrained models, while performing well on many tasks, ``have limitations on certain basic symbolic manipulation tasks such as copy, reverse, and addition''~\cite{Qian2022}, particularly when the complexity of the input increases. This performance degradation in tasks involving longer sequences or repeating symbols suggests inherent limitations in the models' ability to generalize and perform structured reasoning.

Furthermore, a recent paper~\cite{west2024the} introduced the concept of the ``Generative AI Paradox,'' highlighting that while generative models can produce outputs resembling expert-level work, their capabilities ``are not contingent upon—and can therefore exceed—their ability to understand those same types of outputs''~\cite{west2024the}, concluding that they often produce outputs without a true understanding of the underlying concepts. This divergence between generation and comprehension underscores the models' reliance on pattern recognition rather than genuine reasoning, leading to outputs that may be syntactically plausible but semantically incorrect.

These issues pose a significant barrier to the unsupervised generation of instructional content in mathematics, where rigor and correctness are paramount.

\subsection{Addressing AI Limitations Through Constrained Assessment Formats}
Despite the well-documented limitations of generative AI in mathematical reasoning, not all assessment types are equally affected by these challenges. In particular, multiple choice questions offer a unique opportunity to effectively leverage generative AI within a highly structured format that naturally constrains the model's output and simplifies the validation process.

Multiple choice assessments define a finite and predetermined set of response options, enabling greater control over both the content and the evaluation process. While often perceived as focusing on concept recognition, in many STEM disciplines these questions still require students to engage in full problem-solving processes to arrive at the correct answer. The key distinction lies in the format’s bounded nature: students select rather than construct their response, and educators can carefully design distractors that reflect common misconceptions or subtle variations of reasoning. This makes multiple choice an ideal context in which to deploy generative AI, as the output can be validated efficiently by a subject matter expert and integrated into teaching materials with reduced risk of conceptual error.

Recognizing these advantages, we developed a structured approach to using generative AI to assist in the creation of multiple choice assessments in mathematics within the Möbius platform. Our methodology focuses on prompt engineering strategies that interface with the platform’s semantic math engine, effectively speaking the language of mathematics in the generative process. This bounded format provides a practical mechanism for safely applying generative AI: subject matter experts can review AI-generated distractors for plausibility and pedagogical value, ensuring alignment with course objectives. As such, Möbius’ use of multiple choice assessments creates a controlled environment in which generative AI can enhance content creation without compromising mathematical rigor or instructional quality. 

In what follows, we describe the technical architecture of Möbius and its mathematical capabilities, highlighting the features that enable dynamic question generation and semantic validation. We then outline the prompt design strategy used to interact with generative AI models for producing multiple choice alternatives. Following this, we provide a discussion of the validation process, to evaluate the quality and efficiency of the generated assessments. Finally, we conclude by discussing future work and broader applications of generative AI in mathematical education. 

\section{Platform Context: Möbius as an Environment for Algorithmic Question Generation}
To support the discussion of AI-driven question generation, we provide a technical and functional overview of the Möbius platform. Möbius is employed in this study as the environment for developing, evaluating, and deploying multiple choice questions in higher education-level mathematics. In this section, we describe the general platform framework, the structure and flexibility of the multiple choice question type, and the underlying system architecture. 

\subsection{Platform Overview}
Möbius is a cloud-hosted, browser-accessible platform designed for the delivery of STEM instruction and assessment. It operates through a web interface that supports a broad range of interactive features, and accepts a variety of input formats, including symbolic and numeric mathematical responses. A core component of its functionality is the integration of a computer algebra system (CAS), which enables symbolic computation for use in algorithmic question development, question evaluation, and feedback generation.

The platform is designed to support online learning and assessment workflows through automated question grading, feedback, and randomized variable generation. These capabilities allow for the delivery of varied question instances to individual learners, reducing question repetition and supporting formative assessment practices. There are sixteen (16) question types available in Möbius beyond Multiple Choice, including Mathematical Formula, Maple-Graded, Spreadsheet, HTML, and Free-body Diagram. Educators engage with the platform through an authoring interface that allows them to define parameters, write evaluation logic, and control the presentation of mathematical content.

This technical structure provides the foundation for exploring how generative AI models can be used to assist in the creation of multiple choice questions within a robust, computation-enabled system.

\subsection{Multiple Choice Question Type in Möbius}
The Multiple Choice (MC) question type in Möbius allows for the presentation of a question prompt alongside a set of predefined response options. One or more answers can be designated as correct, and response-specific feedback can be configured to guide learners after submission.

\begin{figure}[t]
    \centering
    \includegraphics[width=1.0\linewidth]{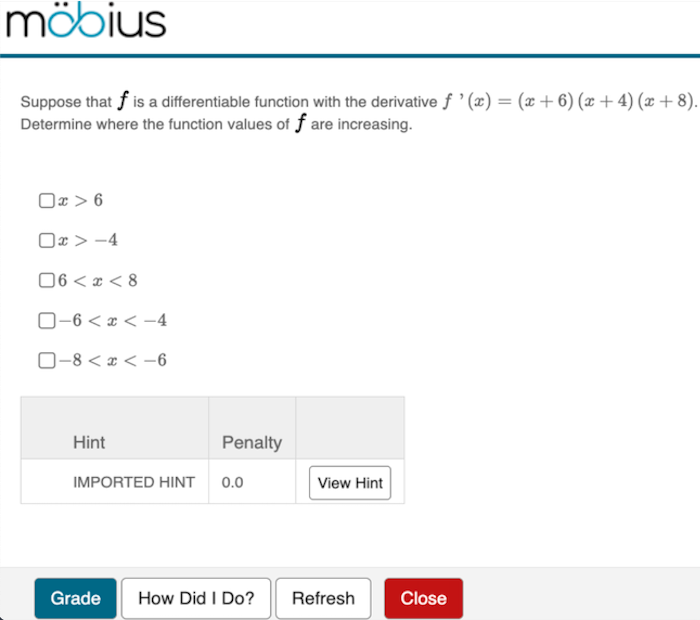}
    \caption{Example of a Multiple Choice question Presented in Möbius.}
    \label{fig:moebiusmc1}
\end{figure}

\begin{figure}[t]
    \centering
    \includegraphics[width=1.0\linewidth]{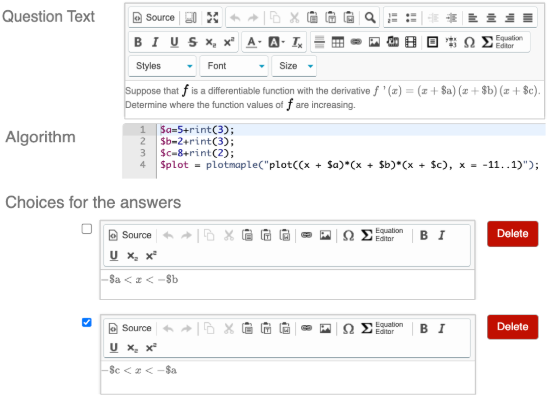}
    \caption{Author's view of Multiple Choice question in Möbius, showing algorithmic variables and an example of symbolic answer options.}
    \label{fig:moebiusmc2}
\end{figure}

Möbius’s implementation of the MC format supports the inclusion of both static and dynamic content. Variables can be randomized using algorithmic definitions, allowing different students to receive different parameterizations of the same question structure. Answer choices can include plain text or mathematical notation using LaTeX or MathML, and each option may itself incorporate randomized expressions or computed values. 
This is shown in Figures~\ref{fig:moebiusmc1} and~\ref{fig:moebiusmc2}.

\vspace{4pt}
Key characteristics include:
\begin{itemize}
    \item Support for dynamic values and variables in prompts and answer options.
    \item Customizable feedback at the level of individual choices.
    \item Support for single- and multiple-answer configurations.
    \item Mathematical notation support for accurate display and preservation of semantics of expressions and equations.
\end{itemize}
\vspace{4pt}
This question format is particularly suitable for integration with generative AI systems due to its structured nature, clear boundaries for response options, semantic understanding of mathematical notation, and capacity to incorporate randomized mathematical content.

\subsection{Möbius System Architecture}
The Möbius platform is built on a layered architecture, organized to separate the presentation of content from the execution of mathematical computations. This structure supports system scalability, modular development, and integration of third-party services. Within the context of using Generative AI to author questions in Möbius, the architecture is as follows:
\vspace{4pt}
\subsubsection{Presentation Layer}
The Presentation Layer is responsible for handling user interaction and orchestrating computational requests. It includes:
\begin{itemize}
    \item \textbf{Möbius UI}: Manages the collection of user input and the delivery of question content using Möbius Algorithm Syntax.
    \item \textbf{Möbius Question Generator}: Parses input from the Möbius UI, substitutes parameters into commands, and directs computation tasks to the appropriate service. It also processes the output from the Services Layer and prepares it for display via the Möbius UI.
\end{itemize}
\vspace{4pt}
\subsubsection{Services Layer}
% With [t], the figure will float to the top of a page and can be referenced as  Fig~\ref{label}
\begin{figure}[t]
    \centering
    \includegraphics[width=0.7\linewidth]{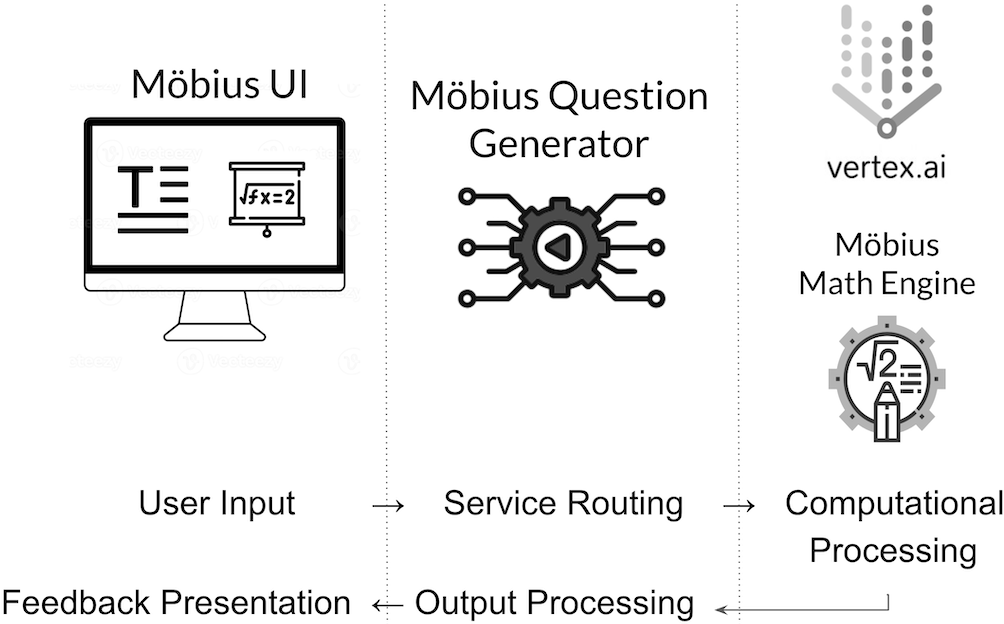}
    \caption{AI-Generated Multiple Choice Questions in Möbius: High-level system diagram.}
    \label{fig:systemarch}
\end{figure}
The Services Layer executes the computational logic needed for question generation and evaluation. It comprises:
\begin{itemize}
    \item \textbf{Vertex AI:} A third-party multi-model AI Machine Learning Engine that executes prompts and instructions from the Möbius Question Generator.
    \item \textbf{Möbius Math Engine:} A core service, orchestrating the execution of algorithmic expressions derived from questions authored in the proprietary Möbius syntax. These expressions can encompass nested code written in Maple, Python, or Java, allowing for complex and multifaceted mathematical computations within a single question.
\end{itemize}
When a user submits a request through the Möbius UI, it hands off the responsibility of routing the input to the appropriate service to the Möbius Question Generator, directing it first to the AI Engine in the case of Prompt responses, then to the Möbius Math Engine which determines whether the native math engine, MapleNet, Python or Java services are appropriate for computation, depending on how the question is authored. The results are returned, interpreted, and presented with feedback through the Möbius UI. This is shown in Figure~\ref{fig:systemarch}.

This structure allows for a combination of real-time symbolic computation, parameterized content rendering, and automated evaluation in a manner that is extensible and adaptable to new question generation techniques.

\section{Designing Effective Prompts for Generating Multiple Choice Alternatives}
\subsection{The Role of Distractors in STEM Multiple Choice Assessment}
Distractors in multiple choice questions serve a substantial pedagogical function by targeting common errors, reinforcing conceptual understanding through contrastive reasoning. In the context of STEM education, distractors are not merely incorrect answers; rather, they represent plausible alternatives that require the student to actively engage with the problem-solving process to identify the correct response. 

In mathematics, distractors often embody systematic errors such as algebraic missteps, sign mistakes, or misunderstandings of function properties. These tailored ``hallucinations'' prompt students to differentiate between conceptual understanding and surface-level familiarity, thus fostering deeper learning. Earlier work~\cite{Haladyna2002} affirms the educational value of well-crafted distractors, noting that they ``can elicit diagnostic information about students misconceptions, which is valuable for instructional decisions''~\cite{Haladyna2002}.

\subsection{Pedagogical Challenges with Distractors}
Despite their instructional benefits, distractors can also introduce pedagogical risks if not carefully constructed. Poorly designed distractors may either be too obviously incorrect, rendering them ineffective, or unintentionally correct, undermining the assessment’s integrity. Further, if distractors reinforce errors without resolution, they may entrench misunderstandings rather than clarify them. According to~\cite{ware2009}, distractors must balance plausibility with pedagogical intent: ``distractors should represent common errors but must be clearly distinguishable from the correct option by students with sufficient understanding of the material.'' This highlights the need for rigorous review and validation of AI-generated distractors, particularly in disciplines like mathematics where logical precision is paramount.

\subsection{Generating Credible Distractors}
In our implementation, we developed a structured prompt strategy to guide the generative AI model in producing high-quality multiple choice alternatives. Key to our success was the use of curated prompt instructions that clearly delineated the intended structure and output format of the responses, from the instructions provided to the AI engine to generate content. Empirical findings~\cite{Qian2022} underscore this strategy, demonstrating that with carefully designed prompts, offering guidance akin to tutoring, AI systems can achieve improved accuracy on mathematical reasoning tasks.

\begin{figure*}[t]
    \centering
    \includegraphics[width=1.0\linewidth]{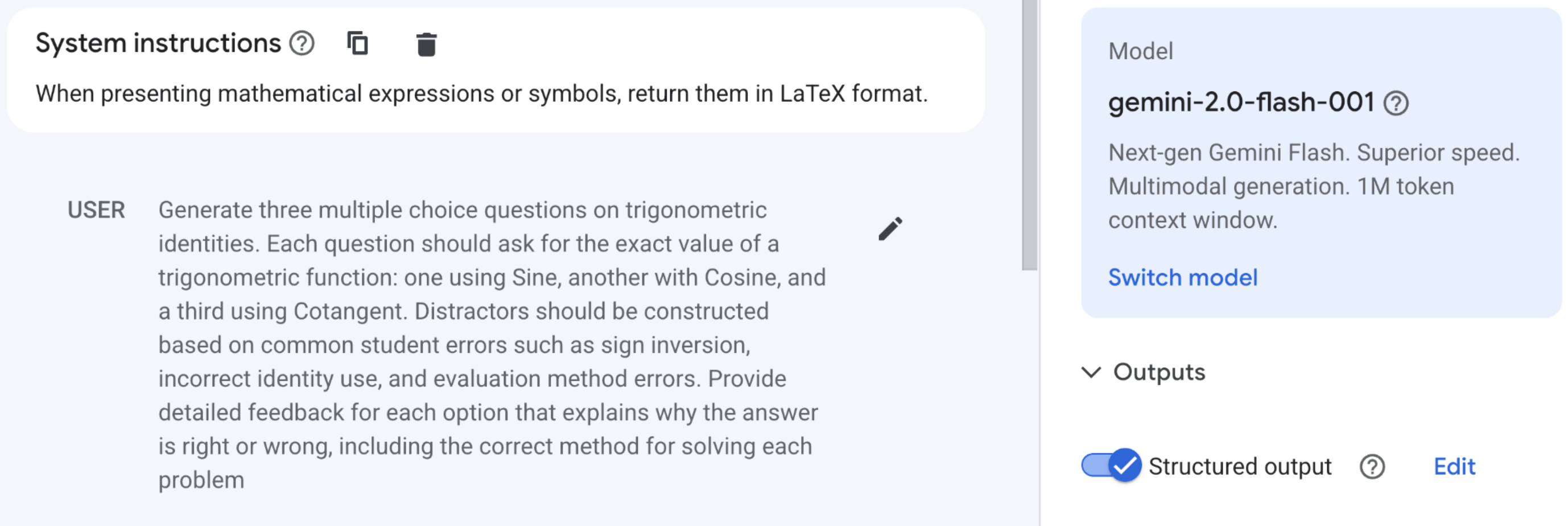}
    \caption{Utilizing the Vertex AI User Interface to demonstrate the structured prompt strategy in practice}
    \label{fig:vertexui}
\end{figure*}

\begin{figure}[t]
    \centering
    \includegraphics[width=1.0\linewidth]{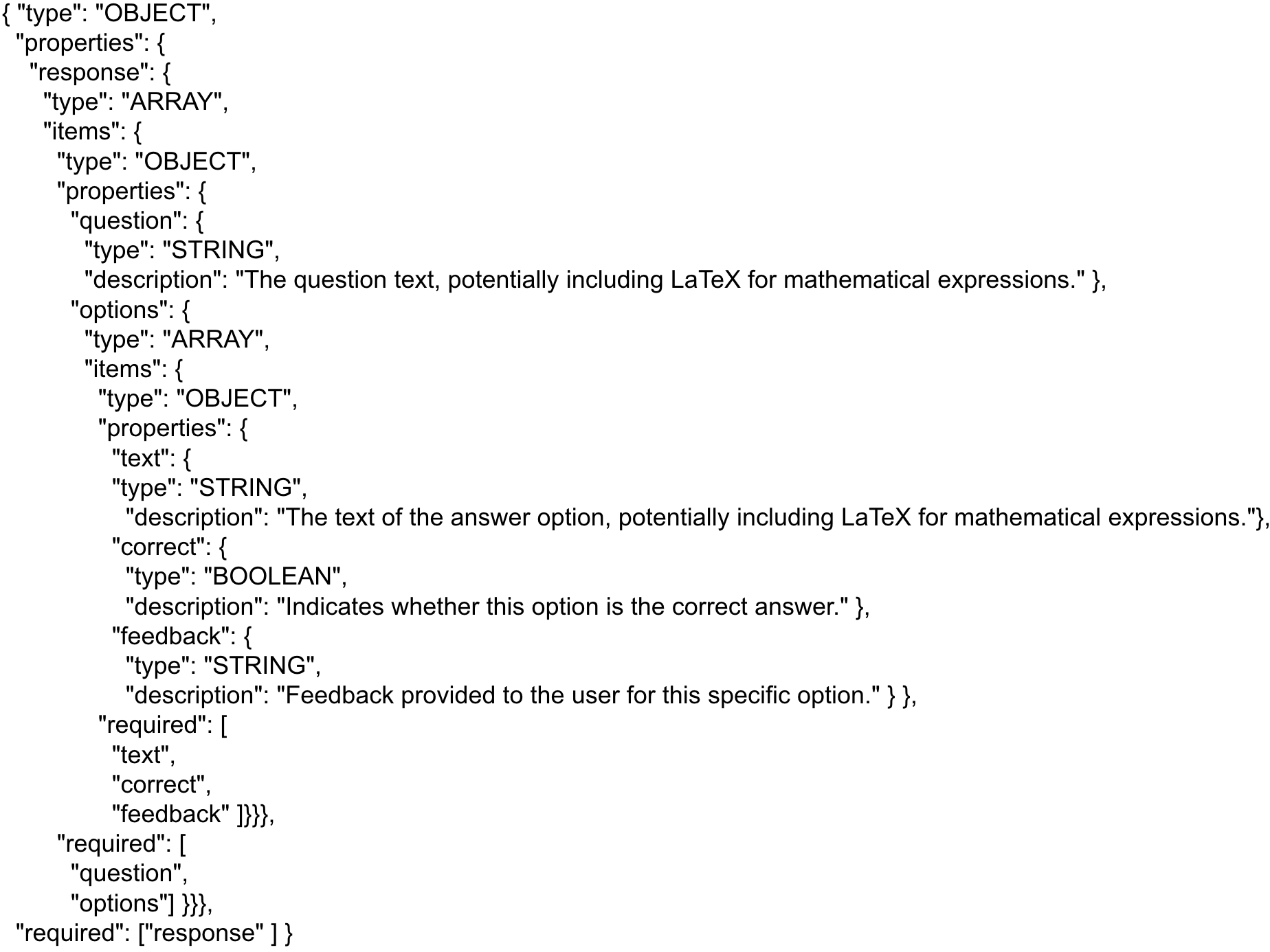}
    \caption{Vertex AI Structured Output Instructions: system instructions provided to the AI Engine regarding the desired response schema}
    \label{fig:schema}
\end{figure}

\begin{figure}[t]
    \centering
    \includegraphics[width=1.0\linewidth]{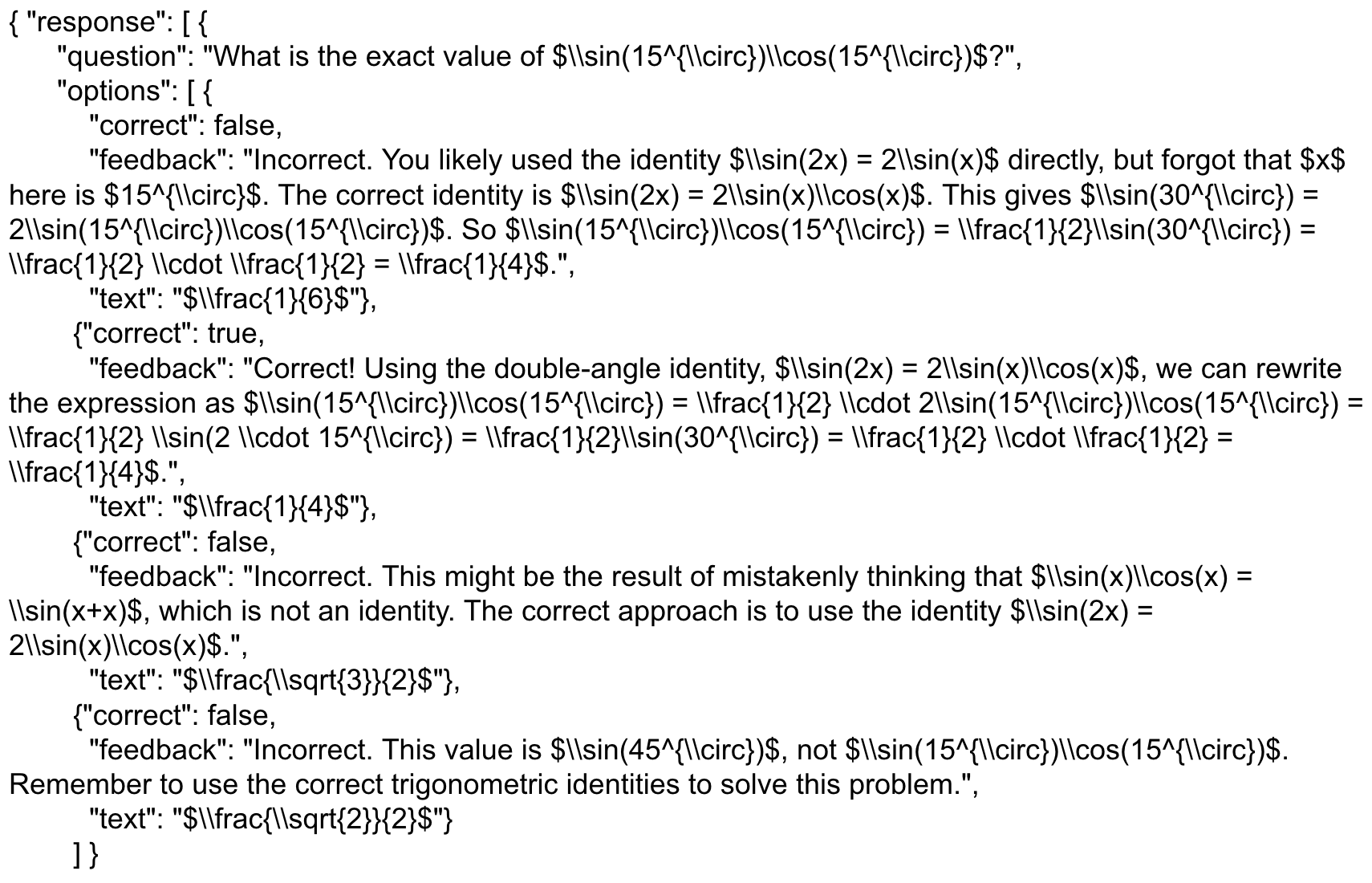}
    \caption{AI-Generated questions and solutions with distractors,  structured in accordance with System Instructions and Response Schema}
    \label{fig:structuredoutput}
\end{figure}

For example, a Subject Matter Expert (SME) designing prompts for a precalculus assignment, might use the following Prompt to generate three multiple choice questions within the domain of trigonometric identities, also shown in Figure~\ref{fig:vertexui}:
\begin{aiprompt}
\textit{Generate three multiple choice questions on trigonometric identities. Each question should ask for the exact value of a trigonometric function: one using Sine, another with Cosine, and a third using Cotangent. Distractors should be constructed based on common student errors such as sign inversion, incorrect identity use, and evaluation method errors. Provide detailed feedback for each option that explains why the answer is right or wrong, including the correct method for solving each problem.}
\end{aiprompt}
Further, knowing that we aim to implement the workflow with the Möbius Math Engine in the loop as part of the overall system architecture, we offer system instructions paired with a schema for the output that specifies the structure and format of the output as part of the input to the AI Engine, as shown in Figures~\ref{fig:vertexui} and~\ref{fig:schema} to ensure that the semantic content of the mathematics is preserved during AI output generation and can be reliably interpreted by Möbius for further processing and validation. An example of the Vertex AI output is shown in Figure~\ref{fig:structuredoutput}. 

%\subsubsection{Tuning the Difficulty of Assessments}
%Difficulty calibration in AI-generated assessments can be achieved by modifying what is presented as options to the student, alongside the correct response. 
%For instance, prompts can specify:
%\begin{aiprompt}
%\textit{Ensure distractors are easily dismissed by students familiar with the concept.}
%\end{aiprompt} 
%to create lower-difficulty questions, or alternatively: 
%\begin{aiprompt}
%\textit{Ensure distractors reflect common stepwise errors such that they appear %plausible without thorough computation.}
%\end{aiprompt}
%to increase difficulty. This capability enables educators to fine-tune questions based on learning objectives, formative vs. summative use, or placement within a curriculum.

\subsection{Verifying the Uniqueness of the Correct Option}
Ensuring that each multiple choice item contains a single, clearly identifiable correct answer is essential for both fairness and validity. 
To support this, we designed a verification process that incorporates the Möbius Math Engine within the Question Generation workflow. At the time of writing, this component of the overall question generation process is architectural, and not fully implemented. 

\begin{figure*}[t]
    \centering
    \includegraphics[width=1.0\linewidth]{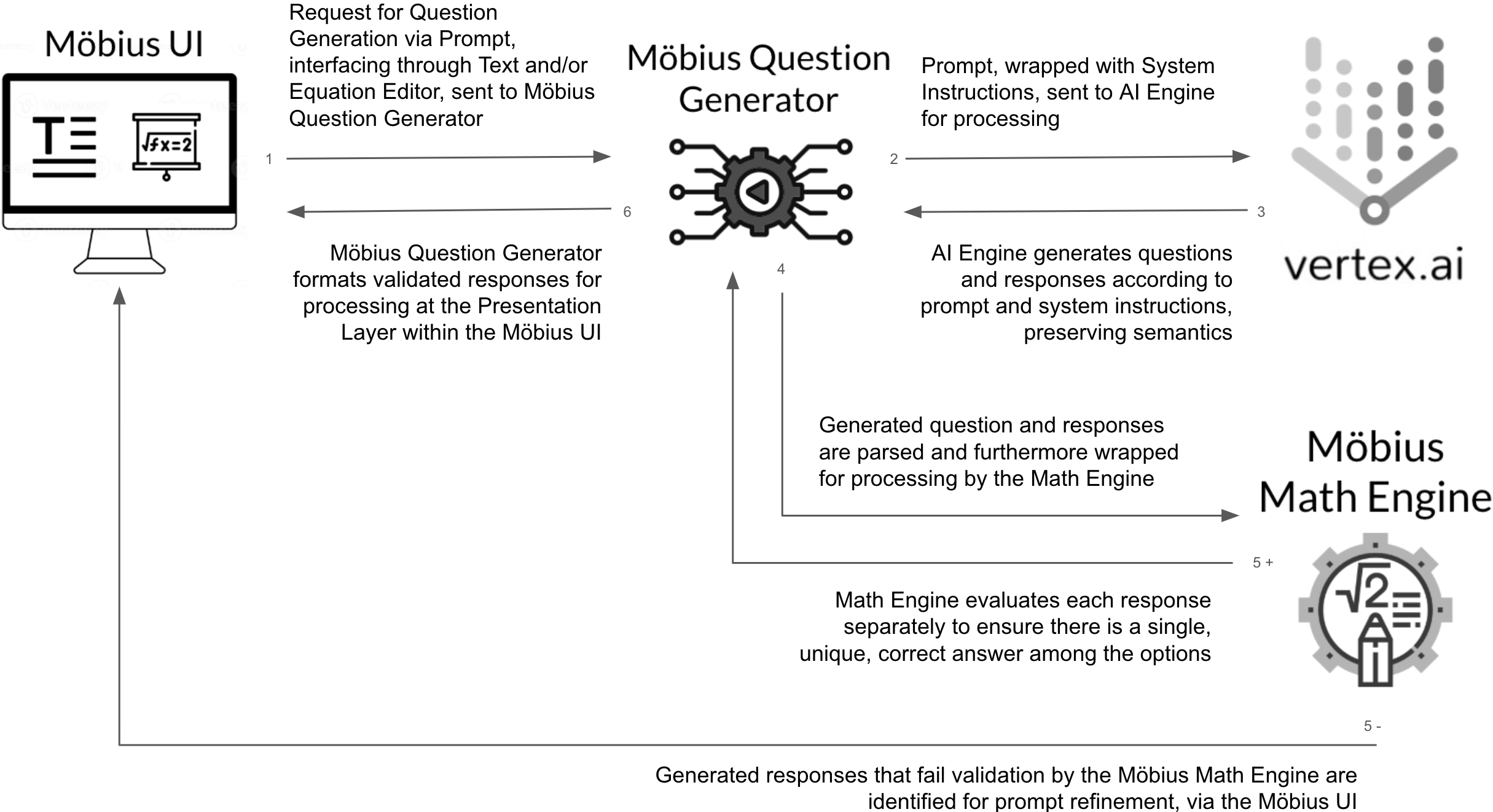}
    \caption{A flow diagram showing the validation process of multiple choice options within the Möbius computation pipeline}
    \label{fig:validation}
\end{figure*}

In this design, after distractors are generated, all answer options are submitted to the engine for symbolic evaluation. If more than one option simplifies to the correct answer or represents a valid form of the same solution, the design allows for the prompt to be adjusted and the distractors regenerated. The designed workflow is shown in Figure~\ref{fig:validation}.

Another verification method involves encoding logical constraints directly into the prompt:
\begin{aiprompt}
\textit{Ensure only one answer evaluates to the correct result; all other responses should reflect common incorrect reasoning paths without coinciding with the correct value.}
\end{aiprompt}
This strategy, encompassing prompt design plus post-processing, offers a robust means to maintain answer uniqueness.

\subsection{Preserving Mathematical Semantics}
In mathematics education, the ability to capture and retain the semantic structure of expressions is essential for ensuring both instructional accuracy and computational fidelity. When integrating generative AI into the development of STEM assessments, maintaining this structure becomes especially important to avoid ambiguity in how mathematical content is represented, interpreted, and evaluated.

To support this, Möbius includes a built-in equation editor that allows authors and students to enter mathematical expressions using traditional two-dimensional notation. This input method mimics how mathematics appears in textbooks or on paper, reducing the likelihood of misinterpretation and ensuring accessibility. The editor encodes expressions in a mathematical markup language that captures not only the visual layout but also the hierarchical meaning of mathematical components. For example, a derivative such as $\frac{d}{dx} \sin{x^2}$, is represented in a way that preserves operator precedence, function nesting, and variable binding, allowing for unambiguous interpretation.

Preserving this semantic detail is particularly useful in workflows that involve large language models (LLMs). When prompts are designed to elicit responses containing structured mathematical content, such as MathML or LaTeX, these formats support the downstream integration of AI-generated material into environments where further processing, such as symbolic manipulation or automated grading, is required.

In Möbius, AI-generated expressions formatted in MathML or LaTeX can be ingested by the platform's question generator and computation engine. This allows for validation of answers, application of algorithmic transformations, and automated feedback generation. Importantly, this workflow reduces reliance on natural language interpretation of mathematical content, which remains a common source of error in large language models when expressions are represented as plain text~\cite{Yamauchi2023}.

By encoding mathematical intent from the point of input and maintaining it through the AI generation and assessment lifecycle, this approach helps ensure that both the instructional and computational goals of mathematics assessments are supported effectively.

\section{Design Validation: Ensuring Reliability and Repeatability}
The integration of generative AI models into structured STEM assessment platforms necessitates robust testing to ensure that the resulting workflows are reliable, repeatable, and pedagogically sound. In the context of Möbius, testing focuses on validating the full authoring pipeline, starting from the initial AI prompt to the delivery and validation of questions within the platform. This section outlines the multi-layered testing approach used to ensure the technical and instructional integrity of AI-assisted question generation.

\subsection{Validating Mathematical Encoding in Prompts and Responses}
Accurate handling of mathematical expressions is foundational to meaningful STEM question generation. In this workflow, authors use the Möbius UI and its Equation Editor to represent mathematical expressions and symbols, which are then translated to a Mathematical Markup Language in prompts submitted to Vertex AI. Since these encodings carry both the visual and semantic meaning of mathematics, any corruption or misinterpretation compromises the integrity of the resulting question.

To confirm the fidelity of mathematical inputs and outputs, we perform testing using representative prompts that include complex expressions, such as nested fractions, function calls (e.g., $ \sin(\frac{\pi}{4})$), and symbolic algebra. Each test ensures:
\begin{itemize}
    \item Prompts containing mathematical expressions and symbols are translated from the Möbius User Interface into the Question Generator, and transmitted in a markup language to Vertex AI without alteration or truncation.
    \item Returned responses containing markup language elements are structurally complete and interpretable.
    \item No semantic distortion occurs due to encoding mismatches or special character handling.
\end{itemize}

This step ensures that mathematical expressions preserve their “intent,” a theme explored in the Preserving Mathematical Semantics section, where we discuss how expression fidelity underpins the platform’s computational accuracy.

\subsection{Verifying JSON Response Parsing and Data Structure Integration}
Vertex AI returns structured data, for Möbius purposes as JSON, which must be parsed and mapped to internal Möbius data structures representing the multiple choice question components. These include the problem stem, options list, feedback for each option, and the identifier for the correct choice.

Validation at this stage focuses on ensuring the structural and functional integrity of the AI-generated response. This includes confirming that the JSON output produced by the model remains consistent when following system instructions, the JSON Schema and that Möbius can reliably parse this structure. Each component, such as the question stem, option text, feedback explanations, and correct answer designation, must be correctly extracted and stored in the appropriate fields for presentation to the user within the Möbius UI. Additionally, LaTeX expressions (or other mathematical markup outputs) embedded in JSON strings are tested to ensure they are safely decoded and accurately rendered within the Equation Editor in Möbius, preserving the mathematical fidelity of the generated content.

Test cases also simulate malformed responses. For instance, missing feedback fields or ambiguous correct answer tags, to ensure that such anomalies result in helpful error handling, rather than silent failure.

\subsection{Simulating Vertex AI API Failures and Error Handling}
To ensure robust operation under real-world conditions, we introduce simulated failure scenarios in communication with the Vertex AI API, and across the Möbius Computation Pipeline. These include:
\begin{itemize}
    \item Submitting prompts with invalid schema or forbidden characters to provoke input validation errors.
    \item Deliberately exceeding API rate limits to test throttling behaviour.
    \item Triggering internal Vertex AI model errors by requesting unsupported output formats or model configurations.
\end{itemize}

Möbius is equipped to detect and manage a variety of error conditions that may arise when interacting with the Vertex AI API, aligning with the broader workflow described in Figure~\ref{fig:validation} for validating and processing AI-generated multiple choice questions. 
When issues such as malformed prompts, model inference errors, or rate limits are encountered, user-facing messages that describe the nature of the error are presented, helping users understand and respond to the issue. Consistent with the system’s emphasis on preserving prompt integrity and mathematical semantics, Möbius also logs relevant API error codes and associated prompt metadata to support debugging and traceability. 

\subsection{Validating AI-Generated Content Through the Möbius Math Engine}
Beyond structural and encoding checks, the final layer of testing is designed to occur within Möbius’s validation framework. The architecture presented is designed to verify the correctness, uniqueness, and evaluability of generated multiple choice content using the Möbius Math Engine, as part of the AI-generated question output. What it is currently not designed to do is emulate Manual Evaluation methods that evaluate plausibility, grammar, diversity of options, and relevance to the context, and described in~\cite{Alhazmi2024}.

This validation process architecture builds upon two earlier concepts: \textit{Verifying Uniqueness of the Correct Answer} and \textit{Preserving Mathematical Semantics}. The key goals at this stage include:
\begin{itemize}
    \item Ensuring that the designated correct answer satisfies the symbolic or numeric evaluation criteria established for the question.
    \item Confirming that each distractor yields a distinct incorrect result when processed by the computation engine, and does not duplicate or approximate the correct answer.
    \item Validating that the feedback logic executes correctly.
\end{itemize}

As described earlier in this paper, the Möbius Math Engine operates on semantically encoded input, allowing it to evaluate answers not just as strings, but as mathematical objects. This supports transformation, simplification, and equivalence checking; a critical capability for STEM assessments where answers can take multiple valid forms.

\begin{figure}[t]
    \centering
    \includegraphics[width=1.0\linewidth]{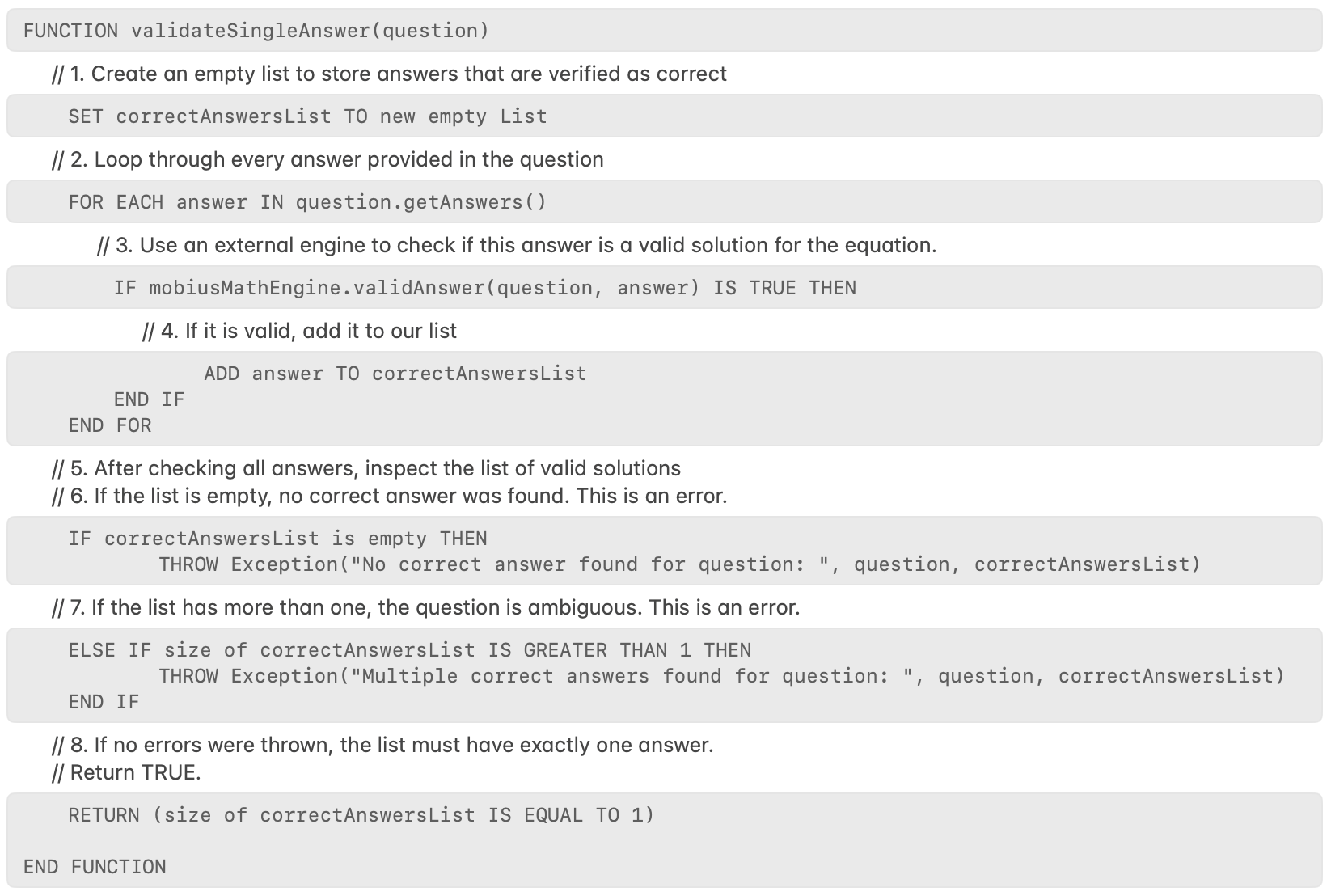}
    \caption{Logic structure for Math Engine evaluation}
    \label{fig:evaluationLogic}
\end{figure}

As illustrated in Figure~\ref{fig:validation}, the validation workflow is designed to trace each option through a parsing, evaluation, and feedback determination sequence, anchored by the logic structure captured in Figure~\ref{fig:evaluationLogic}. 

This logic verifies that each question contains exactly one mathematically correct answer by checking all options against the Möbius Math Engine. If any option fails, it is flagged for revision at the AI-prompt level. This end-to-end process ensures that the resulting question is both pedagogically sound and technically accurate before deployment. 

\balance
\section{Conclusions}
We have explored how generative AI can be used for multiple choice student assessments in mathematical subjects, where structured formats and well-defined concepts mitigate many of the common challenges associated with AI-generated content.  We have done this in the context of the M\"obius platform for online learning.

%By aligning prompt design with the semantic infrastructure of the Möbius platform, 
We have demonstrated that generative models can produce plausible, pedagogically valuable distractors, despite their known limitations in mathematical reasoning and symbolic manipulation.

Our findings suggest that, when used within a constrained framework, generative AI offers meaningful advantages. It
%\begin{itemize}
%\item 
reduces the time and effort required to create high-quality assessments,
%\item 
supports content variation through parameterization, and 
%\item 
retains academic rigour by allowing expert validation at key integration points. 
%\end{itemize}
This approach not only addresses practical concerns in content development, but also underscores a productive reinterpretation of AI ``hallucinations'',  reframing them as opportunities for convincing distractors. As Alhazmi et al.~\cite{Alhazmi2024} note in their survey of distractor generation research, recent advances in large language models have improved contextual coherence but still struggle with producing distractors that are pedagogically valid and semantically distinct. This reinforces our emphasis on validation workflows that preserve mathematical rigor while leveraging AI’s generative potential.

Mathematical subjects are particularly susceptible to a generate-validate workflow.
In many cases, generated alternatives may be checked by a numerical or symbolic mathematical engine.
Future work will extend this work for commercial applications, ensuring that the validation process is robust, transparent, and scalable. While this study relied on a sequenced testing loop within Möbius to verify answer uniqueness, mathematical coherence, and semantic integrity, future work must focus on formalizing and automating this validation pipeline. This includes developing automated test suites for symbolic equivalence, mathematical correctness, and user feedback for prompt refinement. 
% Integrating such validation steps as modular, repeatable components within the question generation workflow, will be essential to ensure trust, reduce review costs, and enable safe reuse of AI-generated content.

% Our findings show 
Finally, we have seen
that generative AI can significantly accelerate the creation of STEM assessments when paired with discipline-specific platforms and rigorous validation. Formalizing the validation layer is the next critical step in moving from proof-of-concept to sustainable, commercial-scale deployment.

\vspace{12pt}
\bibliographystyle{plain}
\IfFileExists{IfExistsUseBBL.tex}{%

}{%
\bibliography{main.bib}
}

\end{document}